\documentstyle[11pt, aaspp4]{article}
\begin{document}
\newcommand{\cc}{\mbox{cm$^{-3}$}}
\newcommand{\tauv}{\mbox{$\tau_V$}}
\newcommand{\ra}{\mbox{$\rightarrow$}}
\newcommand{\nhtwo}{\mbox{n$_{H_{2}}$}}

\def\HI{H{\smc I}}
\def\HII{H{\smc II}}
%                                      molecules
\def\m17{M~17}               % M 17    
\def\cepa{Cepheus~A}               %  Ceph A    
\def\Htwo{H$_2$}               % H2    
\def\HtwoO{H$_2$O}             % H2O 
\def\HtwoeiO{H$_2^{18}$O}             % H2(18)O 
\def\HtwoCO{H$_2$CO}           % H2CO 
\def\HtwoCS{H$_2$CS}           % H2CS 
\def\Hthreep{H$_3^+$}          % H3+
\def\HtwoDp{H$_2$D$^+$}        % H2D+
\def\HthreeOp{H$_3$O$^+$}          % H3+
\def\HCOp{HCO$^+$}             % HCO+
\def\DCOp{DCO$^+$}             % DCO+
\def\HthCOp{H$^{13}$CO$^+$}    % H13CO+
\def\HtwCsiOp{H$^{12}$C$^{16}$O$^+$} % H12C16O+
\def\HCSp{HCS$^+$}             % HCS+
\def\HthCN{H$^{13}$CN}         % H13CN 
\def\HCfiN{HC$^{15}$N}         % HC15N 
\def\HtwCfoN{H$^{12}$C$^{14}$N}  % H12C14N 
\def\HNthC{HN$^{13}$C}         % HN13C
\def\HfoNtwC{H$^{14}$N$^{12}$C}  % H14N12C
\def\HCthreeN{HC$_3$N}         % HC3N 
\def\twCO{$^{12}$CO}           % 12CO
\def\thCO{$^{13}$CO}           % 13CO
\def\CseO{C$^{17}$O}           % C17O
\def\CeiO{C$^{18}$O}           % C18O
\def\twCsiO{$^{12}$C$^{16}$O}  % 12C16O
\def\thCsiO{$^{13}$C$^{16}$O}  % 13C16O
\def\twCeiO{$^{12}$C$^{18}$O}  % 12C18O
\def\thCeiO{$^{13}$C$^{18}$O}  % 13C18O
\def\CtfS{C$^{34}$S}           % C34S
\def\thCS{$^{13}$CS}           % 13CS 
\def\twCttS{$^{12}$C$^{32}$S}  % 12C32S
\def\tfSO{$^{34}$SO}           % 34SO
\def\ttSsiO{$^{32}$S$^{16}$O}  % 32S16O
\def\SOtwo{SO$_2$}             % SO2 
\def\tfSOtwo{$^{34}$SO$_2$}    % 34SO2
\def\SiO{SiO}             % SiO 
\def\Ntwo{N$_2$}               % N2
\def\Otwo{O$_2$}               % O2
\def\NtwoHp{N$_2$H$^+$}        % N2H+ 
\def\NHthree{NH$_{3}$}         % NH3
\def\CHthreeCCH{CH$_3$C$_{2}$H}     % CH3CCH
\def\CHthreeCN{CH$_3$CN}       % CH3CN
\def\CHthreeOH{CH$_3$OH}       % CH3OH
\def\CHfour{CH$_4$}       % CH4
\def\COtwo{CO$_2$}       % C02
\def\thCHthreeOH{$^{13}$CH$_3$OH}       % 13CH3OH
\def\twCHthsiOH{$^{12}$CH$_3$$^{16}$OH} % 12CH316OH
\def\CtwoH{C$_2$H}             % C2H
\def\CtwoS{C$_2$S}             % C2S
\def\CHp{CH$^{+}$}             % CH+
\def\Cp{C$^+$}             % C+
\def\CthreeHtwo{C$_3$H$_2$}    % C3H2
%                                      J transitions
\def\Jthoh{$J = 3/2 \to 1/2$}
\def\Johoh{$J = 1/2 \to 1/2$}
\def\Jtwel{$J = 12 \to 11$}
\def\Jelt{$J = 11 \to 10$}
\def\Jtn{$J = 10 \to 9$}
\def\Jne{$J = 9 \to 8$}
\def\Jes{$J = 8 \to 7$}
\def\Jss{$J = 7 \to 6$}
\def\Jsf{$J = 6 \to 5$}
\def\Jff{$J = 5 \to 4$}
\def\Jft{$J = 4 \to 3$}
\def\Jtt{$J = 3 \to 2$}
\def\Jto{$J = 2 \to 1$}
\def\Joz{$J = 1 \to 0$}
%                                      symbols
\def\WCO{W({\rm CO})}
\def\Wtw{W({\rm ^{12}CO})}
\def\Wth{W({\rm ^{13}CO})}
\def\dv{\Delta v}
\def\dvtw{\Delta v({\rm ^{12}CO})}
\def\dvth{\Delta v({\rm ^{13}CO})}
\def\NCO{N({\rm CO})}
\def\Nth{N({\rm ^{13}CO})}
\def\Ntw{N({\rm ^{12}CO})}
\def\NtwCsiO{N({\rm ^{12}C^{16}O})}
\def\NthCO{N({\rm ^{13}CO})}
\def\NthCsiO{N({\rm ^{13}C^{16}O})}
\def\NtwCeiO{N({\rm ^{12}C^{18}O})}
\def\intCO{\int T_R({\rm CO})dv}
\def\inttwCsiO{\int T_R({\rm ^{12}C^{16}O})dv}
\def\intthCsiO{\int T_R({\rm ^{13}C^{16}O})dv}
\def\inttwCeiO{\int T_R({\rm ^{12}C^{18}O})dv}
\def\NHtwo{N({\rm H_2})}
\def\Wtw{W_{12}}
\def\Wth{W_{13}}
\def\kappanu{\kappa_{\nu}}
\def\phinu{\varphi_{\nu}}
\def\taunu{\tau_{\nu}}
\def\dv{\Delta v}
\def\dvFWHM{\Delta v_{FWHM}}
\def\vLSR{v_{LSR}}
\def\Rsol{R_\odot}
\def\Msol{M_\odot}
\def\MMsol{\ts 10^6\ts M_\odot}
\def\MCO{M_{\rm CO}} 
\def\Mvir{M_{\rm vir}}
\def\TAstar{T^*_A}
\def\TAstartwCO{T^*_A(^{12}{\rm CO})}
\def\TAstarthCO{T^*_A(^{13}{\rm CO})}
\def\TAstarCeiO{T^*_A({\rm C}^{18}{\rm O})}
\def\TRstar{T^*_R}
\def\TexCO{T_{ex}({\rm CO})}
\def\Trms{T_{rms}}
%                                      			units
\def\d{^\circ}
\def\h{^{\rm h}}
\def\mi{^{\rm m}}
\def\s{^{\rm s}}
\def\mum{\ts \mu{\rm m}}
\def\mm{\ts {\rm mm}}
\def\cm{\ts {\rm cm}}
\def\percm{\ts {\rm cm}^{-1}}
\def\m{\ts {\rm m}}
\def\kms{\ts {\rm km\ts s$^{-1}$}}
\def\K{\ts {\rm K}}
\def\Kkms{\ts {\rm K\ts km\ts s^{-1}}}
\def\kHz{\ts {\rm kHz}}
\def\MHz{\ts {\rm MHz}}
\def\GHz{\ts {\rm GHz}}
\def\pc{\ts {\rm pc}}
\def\kpc{\ts {\rm kpc}}
\def\Mpc{\ts {\rm Mpc}}
\def\cmsq{\ts {\rm cm^2}}
\def\pcsq{\ts {\rm pc^2}}
\def\dsq{\ts {\rm deg^2}}
\def\debye{\ts10^{-18}\ts {\rm esu}\ts {\rm cm}}

%                                      			journals
%                                      			mathe
\let\ap=\approx
\let\ts=\thinspace

\newcommand{\six}{\mbox{10$^6$}}
\newcommand{\five}{\mbox{10$^5$}}

\title{FORMATION OF INTERSTELLAR ICES BEHIND SHOCK WAVES}

\author{Edwin A. Bergin$^{1}$, 
David A. Neufeld$^{2}$,
and Gary J. Melnick$^{1}$}

\noindent$^1$ {Harvard-Smithsonian Center for Astrophysics, MS-66, 
60 Garden St., Cambridge, MA 02138; 
ebergin, gmelnick@cfa.harvard.edu}\newline
\noindent$^2$ {Department of Physics and Astronomy, The Johns Hopkins University,
3400 North Charles Street, Baltimore, MD 21218; neufeld@pha.jhu.edu}

\begin{abstract}

We have used a coupled dynamical and chemical model to examine the
chemical changes induced by the passage of an interstellar shock in
well shielded regions.
Using this model we demonstrate
that the formation of \HtwoO\ in a shock will be followed in the post--shock phase
by depletion of the water molecules onto the grain surfaces.
To attempt to discriminate between the creation of ices behind shocks
and their production by means of grain surface chemistry,
we examine the deuterium chemistry 
of water before, during, and after a shock. 
We show that chemical evolution in the post--shock gas can 
account for both the deuterium fractionation and the abundance of \COtwo\ relative
to \HtwoO\ observed in interstellar and cometary ices. 
Given the pervasiveness of shocks and turbulent motions within molecular clouds,
the model presented here offers an
alternate theory to grain surface chemistry for the creation of
ices in the interstellar medium, ices that may ultimately be incorporated into
comets. 

\end{abstract}


\received{6 October 1988}
\accepted{10 November 1998}
%\journalid{337}{15 January 1989}
%\articleid{11}{14}

\slugcomment{To appear in {\em Astrophysical Journal Letters}}

\lefthead{}
\righthead{}

\pagebreak

\section{Introduction}

There is mounting evidence that some interstellar material survives inside
objects in the solar system.  Such evidence is found in the compositional
similarity between cometary ices and inter-planetary grains with those
found in the interstellar medium (Mumma 1997).    
That cometary ices and organics may have supplied the molecular seeds for
prebiotic life on the earth, or even the oceans, leads to the intriguing possibility
that the molecules essential for life might have their
origin in interstellar space (e.g. Chyba \&
Sagan 1997).  
As such it is important to gain an understanding of the 
processes that lead to the formation of ices in the interstellar medium (ISM).  

By far, the most abundant constituent of ices in the interstellar medium,
and also in comets, is H$_2$O.  
Because the low temperature ion--molecule chemistry that is active in dense
regions of the ISM is incapable of reproducing
the observed \HtwoO\ ice observations (Jones \& Williams 1984),
it has been proposed that water--ice
is formed through the hydrogenation of oxygen atoms on the surfaces of 
cold ($T_{d} \sim 10$ K) dust grains (c.f. Tielens \& Hagen 1982).  
%This suggestion is based on several lines of thought.
%First, catalyzed reactions that foster the binding of atoms and molecules
%on grain surfaces are known to occur under terrestrial conditions and
%it is reasonable to assume that solid surfaces may also play a similar
%role in regions of the ISM.  Indeed, it is believed that the most abundant
%molecular species, H$_2$, is formed mainly via grain surface reactions 
%(Hollenbach \& Salpeter 1970) and theory predicts that similar reactions are efficient
%enough to create water-ice (Tielens \& Hagen 1982).  
%Second, in 10 million yr, roughly the lifetime
%of giant molecular clouds that are the sites of star formation,
%low-temperature ion-neutral gas phase
%chemistry alone can deposit water molecules onto grain surface in an abundance
%of only $\sim 10^{-5}$ relative to H$_2$ (Bergin, Langer, \& Goldsmith
%1995); this is nearly an order of magnitude below the observed
%solid H$_2$O abundance 
%and suggests that another mechanism,
%such as grain surface reactions, must play an important role (Jones \& Williams 1984). 
Until recently there have been 
few alternatives to grain surface chemistry to
account for the abundance of ices.
However, we have found that water--ice mantles form quite naturally in a layer of
gas rich in water vapor that has been processed by a shock with a velocity in excess of 10
\kms\ (Bergin, Melnick, \& Neufeld 1998; hereafter BMN).  
The prediction that large quantities of water vapor are produced in shocked gas has recently
gained additional support through the direct observations of strong emission from
gaseous water toward Orion BN--KL (Harwit et al. 1998) using the
{\em Infrared Space Observatory}.  
%The association of this
%emission with the Orion shock is supported by the high excitation of several
%of the water transitions as well as the broad intrinsic widths of these lines.
%Given the pervasiveness of
%shocks and turbulent motions in the ISM, depletion onto grains of shock-produced
%water provides a promising mechanism for the creation of 
%ISM ices that may eventually be incorporated into cometary bodies.

There are several observational constraints that might discriminate
between water-ice mantles created behind shocks and those formed by 
chemical processes in grain mantles. 
In this Letter we examine whether post--shock chemistry can account for the
observed abundances of deuterated water (HDO) and \COtwo\  
%which is found to be enhanced relative to that of \HtwoO\
in the ISM and in comets.
%In this Letter we demonstrate that post-shock chemical
%evolution can reproduce the observed HDO/\HtwoO\ ratios.  We
%also find that a natural by-product of the chemical disequilibrium following a shock
%is the production of CO$_2$ in an abundance similar to that observed in ISM and
%cometary ices.   Thus the creation of interstellar ices behind shocks --
%ices that may be eventually incorporated into cometary ices --
%is a viable alternative or supplement to grain surface chemistry. 

\section{Model}

%We have used the same dynamical and chemical model discussed in greater detail 
%in BMN.  
%Briefly, we use a 3--stage model to 
We use the 3--stage model described in BMN to 
examine the evolution of chemical abundances in pre--shocked (stage 1), shocked
(stage 2), and post--shock gas (stage 3).   
The chemical evolution in each stage is treated independently, except that the final 
composition of the preceding stage is used as the initial composition of the
next.  The first stage is assumed to be initially atomic in composition and has
physical conditions appropriate for 
quiescent gas in molecular cores ($T = 10 - 30$ K; \nhtwo\ = $10^4 - 10^6$ \cc ).
The stage~1 pre--shock chemistry evolves until $t$ = 10$^6$ yr when the gas
is assumed to be shocked.  The stage~2 shock dynamics 
are modeled as an increase in the {\em gas} temperature which scales
with the shock velocity --- peak gas temperatures range
between 400 and $\sim$2000 K
for shock velocities, $v_s$, between 10 and 30 \kms, respectively
(Kaufman \& Neufeld 1996).  The high-temperature chemistry in the shock 
is allowed to evolve for a cooling timescale ($\sim 100$ yr for a 30 \kms\ shock;
see BMN), whereupon 
the third stage commences with a return to quiescent temperature conditions.
%This study examines the compositional changes produced within a shock and the
%manner by which the altered chemistry re-attains equilibrium.

We use the UMIST RATE95
database (Millar, Farquhar, \& Willacy 1997) for all computations
of non-deuterated molecules.  For the deuterium chemistry we have created a smaller
network containing all the important species and reactions that lead to the 
formation and destruction of HDO and OD.
The database was created using reactions taken from the literature
(Croswell \& Dalgarno 1985; Millar, Bennet, \& Herbst 1989; 
Pineau des For\^{e}ts, Roueff, \& Flower 1989;
Rodgers \& Millar 1996).   The principle reaction in
the deuterium chemistry is 
$\rm{H_3^+  +   HD   \leftrightarrow   H_2D^+   +   H_2}$, for which 
we have used the rate coefficients in Caselli et al. (1998).
%%forward reaction  rate is 
%%$k_f$(R1)$ = 1.5 \times 10^{-9}$ and the reverse, 
%%$k_r$(R1)$ = 2.0 \times 10^{-9}(T/300)^{-0.8}e^{-(230/T)}$. 
We have adopted the \Hthreep\ electron recombination
rate and branching ratios given in Sundstrom et al. (1994) and Datz et al. (1995),
respectively.
The \HtwoDp\ recombination rate and branching ratios are from Larsson et al. (1996). 
The deuterium chemical network
was tested against the larger network of Millar, Bennet, \& Herbst (1989) and 
the degree of deuterium fractionation in \HtwoDp , OD, and HDO was found to be
in excellent agreement for the temperature ranges considered here.
As a corollary to the high temperature reactions which form \HtwoO\ in the 
shocked gas (see Kaufman \& Neufeld 1996), we have included similar 
reactions which form HDO.  These reactions and rate coefficients are listed in
Table 1.
%(O  $+$   HD   $\rightarrow$   OD   $+$   H;
%$\rm{OD + H_2 \rightarrow HDO + H}$; and $\rm{OH + HD \rightarrow HDO + H}$; see
%Table 1 for rate coefficients).

We have used the gas-grain adaptation of the UMIST network discussed in
Bergin \& Langer (1997).   For the first and second stages we assume that
molecules are depleting onto bare silicate grains, while in the third stage the molecules
deplete onto a grain mantle dominated by solid \HtwoO .  
This will increase the binding potential for all species by a factor of 1.47.  
We have 
used the measured binding energies of CO, CO$_2$, and \HtwoO\ to \HtwoO\ given
in Sandford \& Allamandola (1990).  
%Deuterated molecules have
%a larger binding energy than their hydrogenated counterparts (Tielens 1983) and
%we have therefore increased the binding energy of all deuterated species relative
%to the hydrogen--bearing molecules.  
%In our models cosmic ray desorption and thermal
%evaporation are the main mechanisms to desorb molecules from the mantle.
The dust temperature, which is critically important for the rate of thermal evaporation,
is assumed to be equivalent to the gas temperature, except in
the shock (stage 2) where we artificially raise the temperature to 200 K to account for
the removal of the ice mantle due to sputtering or grain--grain collisions.

\section{Results}

Figure~1 presents the time evolution of abundances for the 
3--stage gas--grain chemical model.
The peak gas-phase \HtwoO\ abundance is low
([\HtwoO ]/[\Htwo ] $= 3 \times 10^{-7}$) in the pre--shock gas, rises to $\sim 10^{-4}$ 
following the passage of a shock, and depletes 
onto the grain surface at $t \simeq 10^{5}$ yr (for \nhtwo\ = 10$^{5}$ \cc )
in the post--shock gas. 
%Thus, we reproduce the BMN result that abundant water-ice mantles are created
%after a shock heating event.   
In the shock stage we assume $v_s = 20$ \kms ,  which
will heat the gas to $\sim$1000 K; similar results would be found for any shock
velocity between $\sim$10 and 40 \kms .
In the pre-shock quiescent stage we find significant levels of
D--fractionation: at $t =$ \six\ yr we find [HDO]/[\HtwoO ] $\sim
10^{-3}$ in both the gas and solid phases.  
%This agrees with results from the extensive
%network in Millar et al. (1989) for $T_d = 30$ K.
Thus [HDO]/[\HtwoO ] $>$ [HD]/[\Htwo ] $= 2.8 \times 10^{-5}$, 
which is the result of 
low temperatures and low electron abundances favoring
production of \HtwoDp\ relative to \Hthreep .  These enhancements are mirrored
in the daughter products of \HtwoDp , such as HDO.
In the shock itself HDO is released from the grain surface leading
to an increase in its gas phase abundance.  Thus, in stage 2 the HDO abundance
shows little change because 
the fractionation via the ion-molecule chemistry is halted at higher temperatures.
There is some production via the high--T reactions listed in Table~1;
however, these reactions do not increase 
the fractionation significantly. 
%resulting in [HDO]/[\HtwoO ] $= 5 \times 10^{-5}$.
Therefore, the rapid hydrogenation of oxygen is not followed by a similar increase
in the HDO abundance and in the shock [HDO]/[\HtwoO ] $= 5 \times 10^{-5}$.
Pineau des For\^{e}ts et al. (1989) examined the D-chemistry in shocks
and found little change in the water D/H ratio.  As noted by the authors,
their work did not examine the high temperatures required to 
produce abundant \HtwoO\ and therefore did not probe this change. 

In the post--shock gas (stage 3 in Figure~1) the abundance of gaseous HDO is 
approximately constant with a small, but increasing, abundance frozen on the
grain surfaces.  At later times, $t \sim 4 \times 10^{4}$ yr, 
\HtwoO\ begins to deplete onto the grain surfaces, removing
the primary destruction pathway for \Hthreep\ and
\HtwoDp\ 
%(both species react with \HtwoO\ forming H$_3$O$^{+}$ and HD$_2$O$^{+}$
%respectively).  
As the abundance of \HtwoDp\ ions increase, they 
react directly with water forming HDO (along with OD and O), and further increase
the fractionation.  
%Thus, gaseous 
%HDO will form directly from \HtwoO\ in the post--shock layer and these molecules will
%deplete onto the grain surface, enhancing fractionation.  

As shown the top panels of Figure~1,
the abundance of \DCOp\ is quite high with [\DCOp ]/[\HCOp ]
$= 0.002$, but in the shock the abundances of both \DCOp\ and \HCOp\ decline
(due to reactions with \HtwoO ) and fractionation is reversed such that
[\DCOp ]/[\HCOp ] $\sim 2 \times 10^{-5}$ (see also Pineau des For\^{e}ts et al. 1989).
%The decrease in [\DCOp ]/[\HCOp ] in the shock is discussed 
%in Pineau des For\^{e}ts et al. (1989) and is also in agreement with the non--detection 
%of \DCOp\ in the L1157 outflow (Bachiller \& Per\'ez--Guiti\'errez 1997).
%In the post--shock stage the fractionation ratio resets to the quiescent value. 
Thus, \DCOp\ should be an excellent tracer of quiescent gas in
star forming regions, as chemical theory predicts it will be destroyed in an outflow. 

%The abundance of CO in the first two stages remains effectively unchanged (see
%Figure~1) since active CO sublimation at $T_d = 30$~K suppresses the net depletion
In the first two stages active CO sublimation at $T_d = 30$ K suppresses the net CO depletion
onto bare silicate grains.  In stage~3, at $\sim 10^5$
yr, the gaseous CO abundance declines because
%as the result of two effects.
we have assumed that in the third stage the grains are {\em a priori} coated
by water--ice which results in a greater binding energy, resulting in an exponentially
decreasing sublimation rate 
compared to bare silicates (see \S 2).  This leads
a solid CO abundance 
given by [CO]$_{gr}$/[H$_2$] $ = 8 \times 10^{-5}$ at $t = 10^7$ yr.   
Second, on similar timescales
the abundance of OH is enhanced due to the destruction of water molecules.
The reaction of OH with CO will form
CO$_2$, which readily depletes onto grain surfaces.

\section{Discussion}

\subsection{Comparison with Observations}

%Figure 1 demonstrates that the post--shock gas--phase chemistry can quickly
%($< 10^5$ yr) produce a high degree of D--fractionation and a large abundance
%of \COtwo\ ice prior to the depletion of gas phase molecules.   
Observations of ices in the ISM and in comets have yielded limits on the
level of deuterium fractionation in water along with the quantity of \COtwo\
frozen {\em in the water matrix} (see Whittet et al. 
1998).  
One question is whether our model can simultaneously reproduce
{\em both} the observed [\COtwo ]/[\HtwoO ] and [HDO]/[\HtwoO ] ratios. 
Equilibrium models of diffusion-limited 
grain surface chemistry (i.e. chemistry that is limited by the depletion
rate onto the surface) can reproduce high D--fractionation
of water (Tielens 1983) and produce [\COtwo ]/[\HtwoO ] $\sim 0.01 - 10$\% 
(Tielens \& Hagen 1982; Shalabiea, Caselli, \& Herbst 1998).
%although more recent time--dependent models only obtain [\COtwo ]/[\HtwoO ] $\sim 0.01$\%
%We note that there are no published models of grain surface chemistry treating
%(Shalabiea, Caselli, \& Herbst 1998).
%deuterium fractionation at temperatures greater than 10~K.

In Figure 2 we present the comparison of observations with our model predictions 
as a function of time.  To illustrate the
sensitivity of these results to our assumed gas and dust temperature of 30~K,
results for higher (40~K) and lower (25~K) gas and dust temperatures are also shown.
%This figure shows grain surface concentrations relative to \HtwoO\ only in
%the post--shock stage (stage 3).
The small circles (with error bars) depict
HDO and \COtwo\ measurements for Comets Halley, Hyakutake, and Hale--Bopp,
the only comets for which all of these quantities have been determined.
\footnote{Note that while the three comets have similar [HDO]/[\HtwoO ]
ratios, Comet Hale-Bopp shows a larger [\COtwo ]/[\HtwoO ] ratio than
either Halley or Hyakutake.  These differences are significant
relative to the observational uncertainties, but may be 
an artifact of the various Sun-comet distances at which the
measurements were made: ISO observations of
\COtwo\ in Hale--Bopp were obtained when the comet was 2.9 AU from the Sun
while the other cometary measurements were made at a
Sun--comet distance of $\sim 1$ AU.
The lower binding energy of \COtwo, compared to
\HtwoO , could therefore lead to differences in the
out-gassing of these species at larger Sun--comet distances. 
Interestingly, ISO observations of the short period Comet 103P/Hartley~2 at 1 AU from the Sun 
also indicate high \COtwo\ abundances, but this comet does not have a measured D/H ratio
(Crovisier 1998).  }
%Such differences may require
%a degree of nebular processing of interstellar material.}
In the ISM there are no similar complementary measurements; HDO and \HtwoO\ 
have been observed in the gas phase in molecular hot cores, while 
\COtwo\ ices have only been observed along single lines of sight 
towards stars behind molecular clouds or embedded within them.  
We have therefore represented these two independent
measurements for the ISM as a dotted box in Figure 2. 
%which then covers the observed range of 
%relative abundances in the ISM. 
In hot cores, the gas--phase D--fractionation is believed to trace the solid (D/H) ratio 
because the levels of deuterium fractionation are larger than expected for 
pure gas--phase chemistry 
evolving at the observed 
temperatures (T $\sim 100 - 200$ K).  
Thus, the observed HDO and \HtwoO\ are proposed to be ``fossilized'' 
remnants from
a previous low temperature phase, remnants that were frozen on the grains and evaporate
in the hot core (eg. Gensheimer et al. 1996).

The contours of time-dependent abundance ratios for the range of dust temperatures 
provided in Figure~2 nicely encompass the observations.   For this model to be 
applicable to interstellar ices, the initial gas clouds that collapsed and ultimately
formed the molecular hot cores 
must have evolved at $T \sim 25  - 40$ K to account for
the observed fractionation.  
%That such a temperature range applies is supported by
%measurements of quiescent gas in giant cloud cores (Bergin
%et al. 1994).  
For both \COtwo\ and HDO production and where \COtwo\ ices have
been observed,  our model also suggests that the
quiescent clouds that formed hot cores
would have undergone a shock(s) within the past $10^6 - 10^7$ yr.  This timescale is
consistent with the range of inferred shock timescales in molecular clouds
(Reipurth et al. 1998; BMN).
%We note that this process does not require a single shock, but that successive
%stochastic shock episodes could gradually build up a molecular mantle.

Over the past twenty years or more, there has been considerable
debate about whether comets are composed of relative pristine 
interstellar material or whether significant
processing has taken place within the proto--solar nebula (or indeed
whether comets represent a mixture of pristine and processed
material).  The striking similarity between the composition of comets
and that of interstellar ices, and in particular the very similar
level of deuterium fractionation, might suggest that the
degree of processing in the proto--solar nebula is relatively small.  
That the composition of interstellar ices was preserved during their passage into the
outer parts of the proto--solar nebula is perhaps unsurprising given 
the relatively benign physical conditions; in particular, the
typical shock velocity during accretion onto the trans-Neptunian region 
(where it is believed that the comets originally formed; e.g. Safronov 1969)
was only  $\sim 3\,\rm km \, s^{-1}$ (e.g.\ Neufeld \& Hollenbach 1994),
too small to result in significant water production in the gas phase.

\subsection{Sensitivity to Physical Conditions}

Although our model assumes \nhtwo\ $= 10^5$ \cc , the results are similar at
both higher and lower densities because 
most chemical processes show the same dependence on density.
We adopt $A_V = 10$ mag in our models,
which are therefore only applicable in regions where the ultraviolet field is 
heavily attenuated.
Variations in the dust temperature can affect our results, and our model
%In particular we find that post--shock chemistry can only
can only create ice mantles at dust temperatures
greater than 25 K. 
Below 25~K, \Otwo\ remains
frozen on the grain surface in the pre--shock quiescent stage
leaving little free gas--phase O
for shock chemistry to form \HtwoO\ in abundance.
Under these conditions a problem arises in accounting for the high abundance
of water--ice ($\sim 10^{-4}$; Schutte 1998) 
observed in the cold foreground gas towards Elias 16.  
However, this limitation can be mitigated by adopting
a lower density, \nhtwo\ $\leq$ 10$^4$ \cc , as the reduced depletion rate results in 
a smaller and less constraining \Otwo\ abundance on grains.    

One unresolved issue concerns the large abundance of CO ice predicted
in our model, 
which is quite high ($\sim 70$\% relative to \HtwoO ) 
and is larger than observed in the ISM 
or in comets where abundances are generally $< 20$\% (e.g. 
Mumma 1997). 
However, if we consider a single CO molecule approaching a surface with 
nearly equal amounts of CO and \HtwoO ,  it is difficult to imagine that CO will
always be sharing a (stronger) physisorbed bond with \HtwoO\ rather than a (weaker)
bond to a CO molecule.  In laboratory experiments
where CO and \HtwoO\ (with CO/\HtwoO $= 0.5$) are co-deposited onto metal substrate 
at 10 K, the
evaporation rate of amorphous CO is dominated by the vapor pressure of pure
CO and is not affected by the presence  of \HtwoO\ (Kouchi 1990).  
%Some CO does adsorb
%directly on amorphous \HtwoO\ which evaporates at $T \sim 34$ K. 
Although the sublimation of pure CO is not included in our model (we include
sublimation of CO embedded in a water mantle), in reality some co--deposited
CO will evaporate off the grain mantle, provided that the dust temperature exceeds
17 K, the sublimation temperature of pure CO ice.
The large abundance of frozen CO predicted by our model (under the assumption
that CO sublimation is negligible) may therefore be unrealistic and
should be regarded as an upper limit.
%At lower dust temperatures cosmic rays impacts might provide the necessary non--thermal
%heating impulse to remove some CO (Leger et al. 1985).

%Observations of molecular ice absorption features 
%reveal two different components 
%which are believed to be a water dominated polar matrix, with trace \COtwo\ and
%CO, and an apolar phase dominated by CO, \COtwo , \Ntwo , and \Otwo\ (e.g. 
%Ehrenfreund et al. 1997). 
%These components
%represent either two separate mantle compositions along a given line of sight or
%are suggestive of a layered structure to the mantle.
%Because our models produce mantles which are dominated by \HtwoO\
%(with trace \COtwo\  and CO) they
%are only relevant for the creation of the polar matrix.  However,
%as argued above, some CO (and \Otwo\ and \Ntwo ) would exist in the gas phase
%after the formation of the water mantle in the post--shock stage.  In cold
%regions ($T_d < 20$ K) these gas-phase
%molecules could then partially deplete on top of the water mantle producing layered mantle.
%Such a model is more complicated then presented here and is left for future work. 

\section{Conclusion}

In this Letter we have demonstrated that in well shielded regions
the gas phase chemistry after the passage
of a shock is capable of reproducing the observed abundance of frozen water,
the HDO/\HtwoO\ ratio, and the abundance of \COtwo\ ice. 
Thus, the creation of grain mantles behind shock waves is a viable alternate theory to the
production of ices through grain surface chemistry.  We note that this mechanism does not
preclude grain surface chemistry, and indeed it may act as a supplement.
A future extension of our model will be to determine
whether shock models can account for the enhanced abundances of
methanol observed in shocked regions relative to surrounding gas (e.g. 
Bachiller et al.\ 1995). Such an extension of our model might explain the formation of
methanol ices -- an observed constituent of interstellar
grain mantles -- as arising from the
``freeze-out'' of \CHthreeOH\ gas produced by gas phase chemistry
in the warm shocked gas.  
%This would provide an alternative
%explanation to previous suggestions that methanol ice is
%{\it produced} by grain surface chemistry and {\it released} when
%shock waves lead to grain mantle vaporization. 
%Since the high temperature chemistry 
%relevant to the formation of methanol has yet to be characterized 
%reliably, the question of methanol production by gas-phase reactions 
%behind shocks remains an open one.
Other tests could involve an
examination of other proposed outflow chemistry tracers (e.g. SO) and 
an expanded examination of the deuterium chemistry, including species such
as DCN, HDCO, and D$_2$CO.

It has previously been suggested that the high degree of deuterium fractionation
observed in cold cloud cores can explain the observed deuterium fractionation in
cometary ices (e.g. Geiss \& Reeves 1981).  However, our results indicate that shock waves
are a critical ingredient in the production of ices from gas phase chemistry.
Ion--molecule chemistry in  quiescent gas can reproduce the degree of
deuterium fractionation observed in molecular
hot cores; however, it cannot reproduce the observed
abundance of water--ice (Jones \& Williams 1984; Bergin, Langer, \& Goldsmith 1995).
In our model, the abundance of water--ice in the pre--shock gas at 10$^{6}$ yr 
is $< 10^{-5}$, much less than the value estimated
along the line of sight towards Elias 16 ($\sim 10^{-4}$; Schutte 1998).   
At \nhtwo\ = 10$^{5}$ \cc, the gas--phase chemistry could, with 
time, eventually approach the measured abundance.  However, the molecular ice
absorption observations are taken along a lines of sight that likely 
have lower densities,
which makes reproducing the observed abundances more difficult. 
Thus, either grain surface chemistry, creation of ice mantles behind shock waves, or
other unknown theories, are necessary to account for the presence of 
molecular ices.

We are grateful to A. Dalgarno, N. Balakrishnan, and D. Clary for discussions on
high temperature reactions for deuterium--bearing molecules.  E.A.B. is also
grateful to J. Crovisier for discussions of cometary ice abundances. 
E.A.B.  acknowledges support from NASA's SWAS contract to the 
Smithsonian Institution (NAS5-30702); and
D.A.N. acknowledges the support
of the Smithsonian subcontract SV-62005 from the SWAS program.

\begin{deluxetable}{llrll}
\tablenum{1}
\tablecolumns{5}
\tablecaption{High Temperature Isotopic Reactions\tablenotemark{a}}
\tablehead{
\colhead{Reaction} &
\colhead{A (cm$^{3}$s$^{-1}$)} &
\colhead{B} &
\colhead{C (K)} &
\colhead{Refs} 
}
\startdata
 O  + HD $\rightarrow$ OD  +  H & 1.57 $\times 10^{-12}$ &   1.7 &4639 & 1 \nl
 O  + HD $\rightarrow$ OH  +  D & 9.01 $\times 10^{-13}$ &   1.9 &3730 & 1 \nl
OH  + HD $\rightarrow$ H$_2$O  +  D & 2.12 $\times 10^{-13}$ &   2.7 &1258 & 2 \nl
OH  + HD $\rightarrow$ HDO  +  H & 0.60 $\times 10^{-13}$ &   1.9 &1258 & 2 \nl
OD  + H$_2$ $\rightarrow$ HDO  +  H & 1.55 $\times 10^{-12}$ &   1.6 &1663 & 3 \nl
\enddata
\tablenotetext{a}{Rate coefficient $\rm{k = A(T/300\;K)^{B}e^{-C/T}}$.}
\tablenotetext{1}{Fit to theoretical predictions of Joseph, Truhlar, \& Garret (1988).}
\tablenotetext{2}{Talukdar et al. (1996)}
\tablenotetext{3}{Rate same as OH $+$ H$_2$; Talukdar et al. (1996).}
\end{deluxetable}

\clearpage
\begin{figure}
\figurenum{1}
\plotfiddle{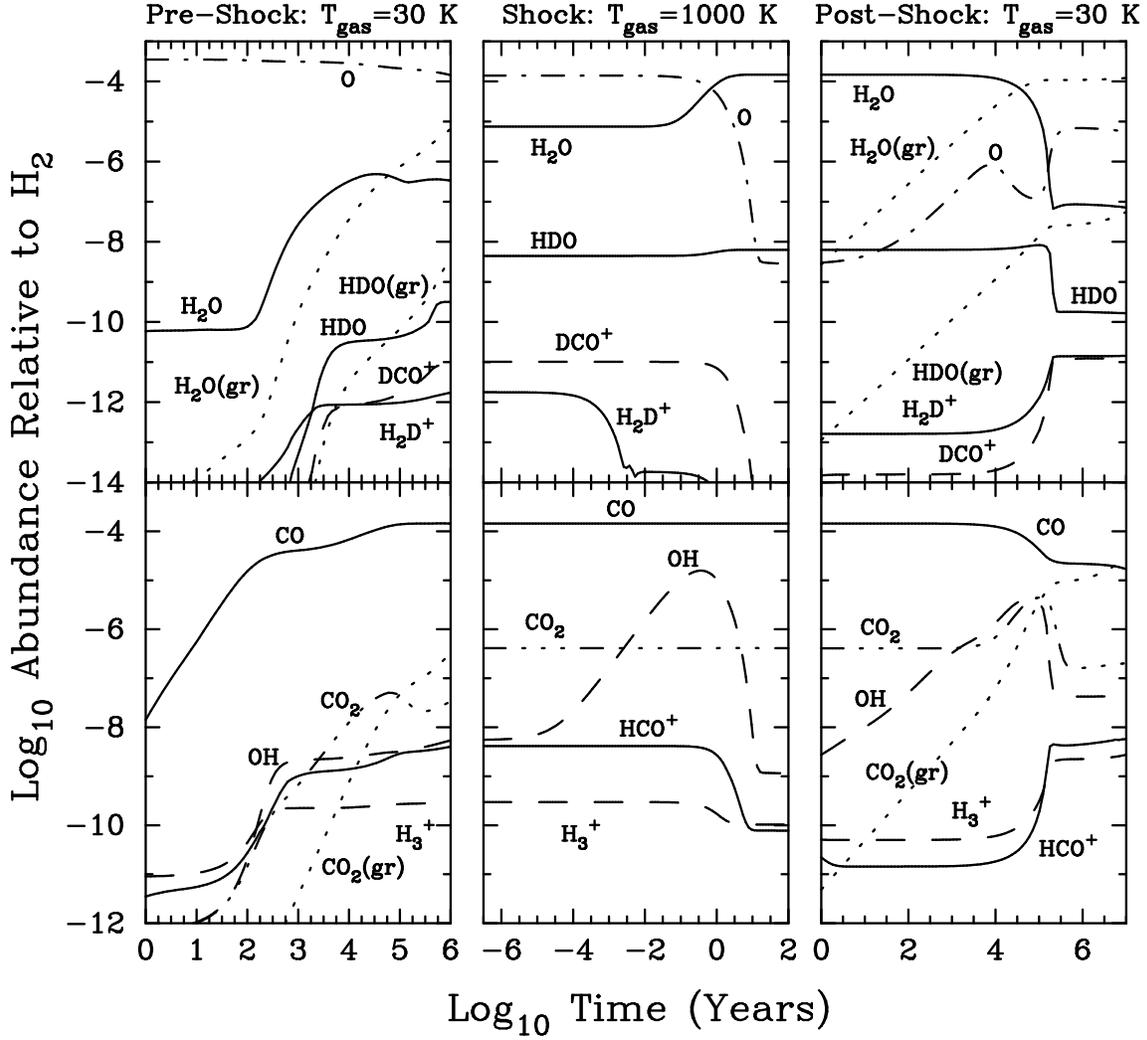}{5in}{0}{80}{80}{-250}{-150}
\label{fig1}
\caption{
Evolution of chemical abundances in our gas-grain three-stage model
as a function of time.  All abundances are relative to H$_2$.
This model includes molecular depletion and desorption from grain surfaces and
the abundance of H$_2$O, HDO, and CO$_2$ on the grain surface is shown.
Stage 1 (first panel) represents the pre-shock stage with $T_{gas} = 30$ K,
Stage 2 (second panel) is the shock stage
with a higher gas temperature ($T_{gas} = 1000$ K; $v_s = 20$ km s$^{-1}$), 
and Stage 3 (third panel) the post-shock stage with $T_{gas} = 30$ K.
The density and visual extinction is constant for all stages at 
n(H$_2$) $= 10^5$ cm$^{-3}$ and $A_V = 10$ mag.
Note that the time axis for
each panel covers a different range and shows the full evolution of
chemical abundances from one stage to the next.
}
\label{fig1}
\end{figure}

\begin{figure}
\figurenum{2}
\plotfiddle{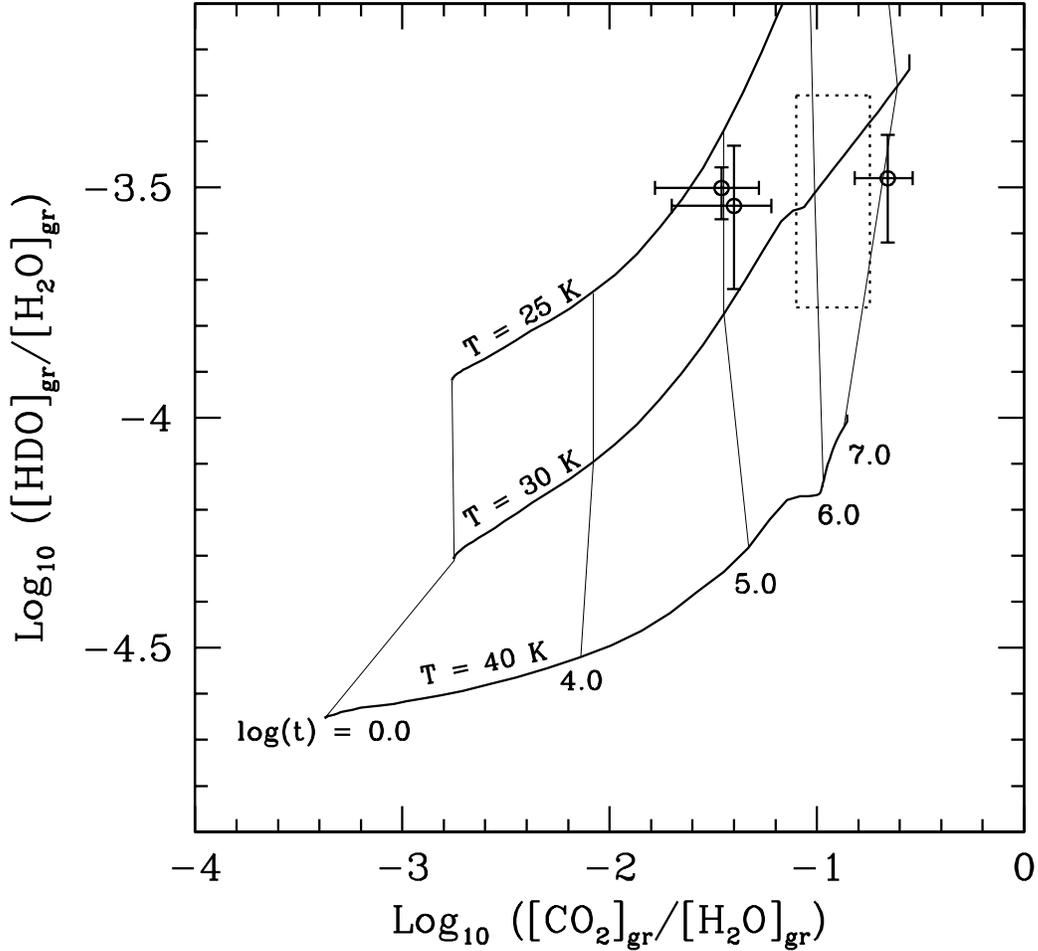}{4.5in}{0}{85}{85}{-300}{-200}
\caption{
Plot of the grain surface [HDO]/[\HtwoO ] against [CO$_2$]/[H$_2$O]  shown
as a function of time [log(t)] and gas and dust temperature (T).
The calculated  abundance ratios are shown only in the post--shock phase (stage 3).  
The points are the observed ratios in Comets Halley, Hyakutake, and Hale--Bopp.
For Halley, Hyakutake, and Hale--Bopp the HDO data are from: 
Eberhardt et al. (1995), Balsiger et al. (1995), Bockel\'ee--Morvan et al. (1998), 
and Meier et al. (1998), respectively.  The CO$_2$ measurements are from 
Kranowsky (1986), McPhate et al. (1996), and Crovisier et al. (1997).
The box shows the range of representative values observed in the interstellar
medium and [HDO]/[\HtwoO ]  ratios are from Gensheimer et al. (1996) 
and the [CO$_2$]/[H$_2$O]
ratios in polar ices are from Schuttte (1998) and Whittet et al. (1998). 
}
\label{fig2}
\end{figure}


\begin{thebibliography}{}

%\bibitem[A'Hearn et al. 1995]{AMSOB95} A'Hearn, M.F.,  Millis, R.L., Schleicher, D.G., 
%Osip, D.J., Birch, P.V. 1995, Icarus, 118, 223

\bibitem[Bachiller \& P\'{e}rez Guti\'{e}rrez 1997]{Bachiller_Perez97} Bachiller, R. \&
P\'{e}rez Guti\'{e}rrez, M. 1997, ApJ, 487L, 93

\bibitem[Bachiller et al. 1995]{BLWC95} Bachiller, R., Liechti, S., Walmsley, C. M., \& Colomer, F. 1995, A\&A, L51
 
\bibitem[Balsiger et al. 1995]{BAG95}
Balsiger, H., Altwegg, K., \& Geiss, J. 1995, J. Geophys. Res., 100, 5827

\bibitem[Bergin, Melnick, \& Neufeld 1998]{BMN98} Bergin, E.A., Melnick, G.J., 
\& Neufeld, D.A. 1998, ApJ, 499, 777

\bibitem[Bergin \& Langer 1997]{BL97} Bergin, E. A., \& Langer, W. D. 1997, 486, 316

\bibitem[Bergin, Langer, \& Goldsmith 1995]{BLG95} Bergin, E. A., Langer, W. D., \& Goldsmith, P. F. 1995, ApJ, 441, 222

%\bibitem[Bergin et al. 1994]{BGSU94} Bergin, E.A., Goldsmith, P.F., Snell, R.L., \&
%Ungerechts, H. 1994, ApJ, 431, 674
 
\bibitem[Bockel\'ee-Morvan et al. 1998]{DBetal98}
Bockel\'ee-Morvan, D.,  et al.
Icarus, 133, 147 

\bibitem[Caselli et al. 1998]{PCetal98} Caselli, P., Walmsley, C.M., Terzieva, R.,
\& Herbst, E. 1998, ApJ, 499, 234

%\bibitem[Caselli, Hartquist, \& Havnes 1997]{CHH97} Caselli, P., Hartquist, T. W., \& Havnes
%, O. 1997, A\&A, 322, 296 
 
\bibitem[Chyba \& Sagan 1997]{CS97} Chyba, C.F. \& Sagan, C. 1997, 
in ``Comets and the Origin and Evolution of Life'', eds. P.J. Thomas,
C.J. Chyba, \& C.P. McKay (New York; Springer), 147

%\bibitem[Clary 1992]{Clary92} Clary, D.C. 1992, J. Chem. Phys., 96, 3656

\bibitem[Croswell \& Dalgarno 1985]{CD85} Croswell, K. \& Dalgarno, A. 1985, ApJ, 289, 618

\bibitem[Crovisier 1998]{Crovisier98} Crovisier, J. 1998, priv. comm. 

\bibitem[Crovisier et al. 1997]{Crovisier97} Crovisier, J. et al. 1997, Science, 275, 1904

\bibitem[Datz et al. 1995]{Datz95} Datz, S., et al. 1995, Phys. Rev. Lett., 74, 896

\bibitem[Eberhardt et al. 1995]{ERKH95} Eberhardt, P., Reber, D., Kranowsky, D., \&
Hodges, R.R. 1995, A\&A, 302, 301

%\bibitem[Ehrenfreund et al. 1997]{PEetal97} Ehrenfreund, P., D'Hendecourt, L., 
%Dartois, E., Jourdain de Muizon, M., Breitfellner, M., Puget, J.L., \& Habing, H.J.
%1997, Icarus, 130, 1

\bibitem[Geiss \& Reeves 1981]{GR81} Geiss, J. \& Reeves, H. 1981, A\&A, 93, 189

\bibitem[Gensheimer, Mauersberger, \& Wilson 1996]{GMW96} Gensheimer, P.D., 
Mauersberger, R., \& Wilson, T.L. 1996, A\&A, 314, 281

\bibitem[Harwit et al. 1998]{HNMK98} Harwit, M., Neufeld, D.A., Melnick, G.J.,
\& Kaufman, M.J. 1998, ApJ, 497, L105

\bibitem[Joseph et al. 1988]{Joseph88} Joseph, T., Truhlar, D.G., \& Barrett, B.C. 1988, J. Chem. Phys., 88, 6982

\bibitem[Jones \& Williams 1984]{JW82} Jones, A.P. \& Williams, D.A. 1984,
MNRAS, 209, 955

\bibitem[Kaufmann \& Neufeld 1996]{KN96} Kaufmann, M. J., \& Neufeld, D. A. 1996, 456, 611

\bibitem[Kranowsky 1991]{Kranowsky91} Kranowsky, D. 1991, in Comets in the Post--Halley
Era, eds. R.L. Newburn, Jr., M. Neugebauer, \& J. Rahe (Dordrecht: Kluwer), 831

\bibitem[Kouchi 1990]{K90} Kouchi, A. 1990, J. Crystal Growth, 99, 1220

\bibitem[Larsson et al. 1996]{Larsson96} Larsson, M., et al. 1996, A\&A, 309, L1

%\bibitem[Leger, Jura, \& Omont 1985]{LJO85} Leger, A., Jura, M., \& Omont, A. 1985,
%A\&A, 144, 147

\bibitem[McPhate et al. 1996]{McPhate96} McPhate, J.B., Feldman, P.D., Weaver, H.A.,
A'Hearn, M.F., Tozzi, G.-P., \& Festou, M.C. 1996, DPS, 28, 09.29

\bibitem[Meier et al. 1998]{Meier98} Meier et al. 1998, Science, 279, 842

\bibitem[Millar et al. 1989]{MBH89} Millar, T.J., Bennet, A., \& Herbst, E. 1989,
ApJ, 340, 906

\bibitem[Millar et al. 1997]{MFW97}  Millar, T. J., Farquhar, P. R. A., \& Willacy, K.
1997, A\&AS, 121, 139
 
\bibitem[Mumma 1997]{Mumma97}  Mumma, M.J. 1997, 
in ``From Stardust to Planetesimals'', eds. Y. Pendleton \& A.G.G.M.
Tielens, (San Fransicso: Ast. Soc. of Pac. Conf. Series), 122, 369

\bibitem[Neufeld \& Hollenbach 1994]{NH94} Neufeld, D.A. \& Hollenbach, D.J. 1994, ApJ,
428, 170
 
\bibitem[Pineau des For\^ets, Roueff, \& Flower 1989]{PdFRF89} 
Pineau des For\^ets, G, Roueff, E., \& Flower, D.R. 1989, MNRAS, 240, 167

\bibitem[Reipurth, Bally, \& Devine 1997]{RBD97} Reipurth, B., Bally, J.,
\& Devine, D. 1997, AJ, 114, 2708

\bibitem[Rodgers \& Millar 1996]{RM96} Rodgers, S.D. \& Millar, T.J. 1996,
MNRAS, 280, 1046

\bibitem[Safronov 1969]{Safronov69} Safronov, V.S. 1969, Evolution of the 
Protoplanetary Cloud and Formation of the Earth and Planets (Moscow: Nauka Press),
[in Russian: english translation NASA TTF-667, 1972]

\bibitem[Sandford \& Allamandola 1990]{SA90} Sandford, S.A. \& Allamandola, L.J.
1990, Icarus, 87

%\bibitem[Schilke et al. 1995]{SKlBPdFR95} Schilke, P., Keene, J., Le Bourlot, J, Pineau des
%For\^ets, G., \& Roueff, E. 1996, A\&A, 294L, 17
 
\bibitem[Schutte 1998]{Schutte98} Schutte, W. 1998, in Laboratory Astrophysics
and Space Research, eds. P. Ehrenfreund \& H. Kochan (Dordrecht: Kluwer), in press

\bibitem[Shalabiea, Caselli, \& Herbst 1998]{SCH98} Shalabiea, O.M., Caselli, P., 
\& Herbst, E. 1998, ApJ, 502, 652

\bibitem[Sundstr\"{o}m et al. 1994]{Sundstrometal94} Sundstr\"{o}m et al. 1994, Science, 263 , 785

\bibitem[Talukdar et al. 1996]{Talukdar96} Talukdar, R.K. et al. 1996, J. Phys. Chem.,
100, 3037
 
\bibitem[Tielens 1983]{T83} Tielens, A.G.G.M. 1983,
A\&A, 119, 177

\bibitem[Tielens \& Hagen 1982]{TH82} Tielens, A.G.G.M. \& Hagen, W. 1982,
A\&A, 114, 245

%\bibitem[Turner 1990]{Turner90} Turner, B.E. 1990, ApJ, 362, L29

%\bibitem[van Dishoeck \& Blake 1998]{vDB98} van Dishoeck, E.F., \& Blake, G.A.
%1998, ARA\&A, 36

\bibitem[Whittet et al. 1998]{Whittet98} Whittet, D.C.B. et al. 1998, ApJ, 498, L159

%\bibitem[Whittet 1993]{Whittet93} Whittet, D. C. B. 1993, in Dust and Chemistry in Astronomy
%, eds. T. J. Millar \& D. A. Williams (Bristol: Institute of Physics), 9
 
%\bibitem[Zhang et al. 1995]{Zhang95} Zhang, D.H., Zhang, J.Z.H., Zhang, Y., 
%Wang, D., \& Zhang, Q. 1995, J. Chem. Phys. 102, 7400

\end{thebibliography}
\end{document}